\documentstyle[prl,aps,multicol,epsf]{revtex}
\draft
\begin{document}
\title{Spectroscopy, Interactions, and Level Splittings
in Au Nano-particles.}
\author{D. Davidovi\'c and M. Tinkham}
\address{Department of Physics and Division of Engineering and 
Applied 
Sciences
\newline Harvard University, Cambridge, MA 02138}
\date{\today}
\maketitle

\begin{abstract}
We have measured the electronic energy spectra of nm-scale Au
particles using a new tunneling spectroscopy configuration.  The 
particle
diameters ranged from 5nm to 9nm, and at
low energies the spectrum is
discrete, as expected by the
electron-in-a-box model. 
The density of tunneling resonances increases rapidly with 
energy, and at higher energies the resonances overlap 
forming broad resonances.
Near the Thouless energy, the broad resonances merge into a continuum.
The tunneling resonances display Zeeman splitting in a magnetic 
field. 
Surprisingly, the $g$-factors ($\sim 0.3$) of energy levels in Au
nano-particles are much smaller than 
the g-factor (2.1) in bulk gold. 
\end{abstract}
\pacs{73.23.-b,73.23.Hk,73.50.-h}

\begin{multicols}{2}
\narrowtext
In nanoscale metallic particles, the electronic energy levels are
quantized due to spatial confinement.  Prior to the present work, this
effect has been studied only by Ralph {\it et al.}
\cite{ralph1}, in small grains of Al.  In this paper, we report
tunneling data on energy spectra in Au nanoparticles, measured with
a new tunneling configuration.
We first show data which qualitatively 
confirm the results of Ralph {\it et al.}
on the Al excitation spectra, but in Au particles. Next, we
confirm the role of the Thouless energy in the spectra of metallic
particles.  We also present new results showing Zeeman splitting 
of energy levels analogous to the Zeeman
effect in Al
particles, but with g-factors much smaller than two. 

To be able to resolve the discrete energy levels in Au at 
$\sim$\,100\,mK, the particle
radius has to be less than approximately 10\,nm.  Fig.~1-a explains 
the steps in our sample fabrication
process. The first step is electron beam lithography. Using a PMMA 
bilayer
resist technique, we define a resist bridge placed
$\sim$200\,nm above the Si wafer; this bridge acts as a mask. 
In step 2, we evaporate
a 15\,nm thick film of Al, along the direction indicated by the
arrow. Then we oxidize the surface of the film in 50\,mtorr of oxygen
for 90\,sec. In step 3, we deposit a 1\,nm thick film of Au, along the
same direction as in step 2. At this stage of film formation, Au forms
isolated particles, with a typical center-to-center spacing of 12\,nm
and a typical base diameter of 6\,nm.\cite{dragoAPL} Next we rotate 
the
sample by $\sim$90$^{\circ}$.  In step 4, we deposit the top 
electrode,
along the direction indicated by the arrow. The top electrode is a
bilayer, which has a $\sim$3\,nm thick layer of Al$_{2}$O$_{3}$ at the
bottom, and a 15\,nm thick film of pure Al on top. The deposition 
angle is
chosen so that the overlap between the bottom and the top electrode is
approximately 20\,nm.  Typically, we grow many electrode pairs
simultaneously, and vary the overlap from 0 to $\sim$30\,nm. Fig.~1-b
is a schematic of the sample at a much larger magnification, showing 
that 
the particle is well screened by the electrodes.  
Because the oxide
between the two electrodes is thick, current between the leads flows
only through the particle(s), due to the exponential dependence of 
tunnel current on barrier thickness. 

\begin{figure}
\epsfxsize=8.6cm
\epsffile{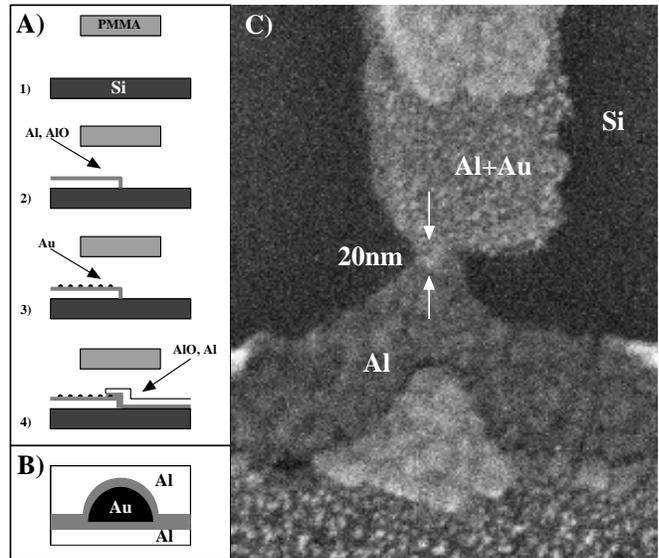}
\caption{a) Fabrication steps of 
our samples. b) Enlarged sketch of the particle. 
c) SEM image of a sample, top view.}
\label{figure1}
\end{figure}

Fig.~1-c is an SEM image of a typical device.  Au nanoparticles are
visible on the top aluminum electrode and on the lower electrode far
from the overlap region. 
In this work, we present data on particles with base diameter
less than 10\,nm, where their shape is roughly hemispherical.
The area of the electrode overlap is such that in most cases, 0-3
particles are covered by the overlap.  From the data, we can determine
whether the tunneling current flows via only one particle, as
explained in the next two paragraphs.
The yield of single-particle samples is about 30\%.

Fig.~2-a displays the I-V curve in sample 1 at 4.2\,K. At
this temperature, the energy levels are not resolved, and at
first sight, the I-V curve is piece-wise linear. The electron
transport at 4.2K is well described by the theory of single charge
tunneling and the Coulomb staircase 
for tunneling via a single particle.\cite{Likharev} The junction 
capacitances $C_{1}$ 
and $C_{2}$, and the 
background charge $q_{0}$ can be determined from
the points where the 
I-V curve changes slope. From the theory\cite{Likharev}, 
we can also estimate the junction resistances from the data. 
A summary of the parameters for 
this and two other samples are shown in Table 1. The
capacitance per unit area in our junctions is $\approx 50fF/\mu
m^{2}$.\cite{jia} 
We estimate the particle base diameter $D$ and volume 
by assuming that the total
capacitance of the hemispherical particle is equal to $C_{1}+C_{2}$.

\begin{figure}
\epsfxsize=9cm
\epsfclipon
\epsffile{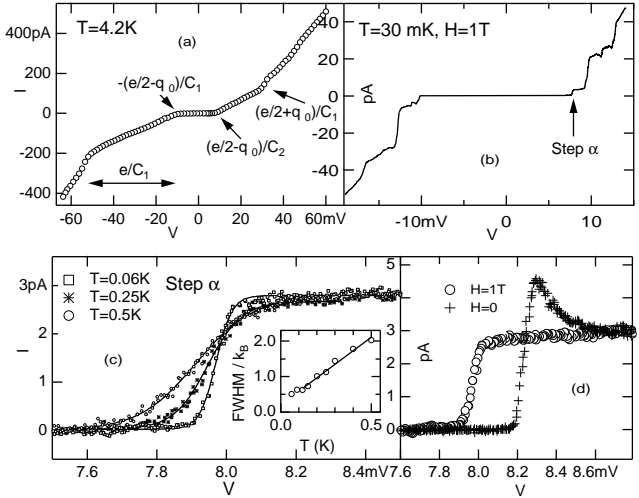}
\caption{a) I-V curve of a 
typical device at 4.2K. 
b). I-V curve of the device at 30mK. c) First current step at three 
different refrigerator temperatures. The inset shows 
full-width-half-maximum 
vs. temperature of step $\alpha$ (after correcting for the capacitive 
division). 
d) Lineshape of the level when the leads are superconducting (H=0) 
compared to the lineshape when the leads are normal (H=1T).}
\label{figure2}
\end{figure}

We apply a number of consistency checks to our measurements 
to ensure
that we are measuring a single particle.
These consistency
checks were introduced and explained in the work by Ralph et
al\cite{ralph1}.  Fig.~2-b presents the I-V curve of the sample at
30\,mK refrigerator temperature on an expanded scale. At low voltages,
current increases in discrete steps. Fig.~2-c shows step $\alpha$ 
in detail
at three different refrigerator temperatures. The curves between the 
points are
fits to the Fermi distribution. The strong temperature dependence of
the line-shape demonstrates that tunneling is via a discrete quantum
level in the particle. After correcting for the 
capacitive division of voltage, the full-width-half-maximum (FWHM)
of the peak is linear with the refrigerator temperature, as shown in 
Fig. 2-c,
with the slope of $3.8k_{B}T$, which is close to the expected slope 
of $3.5k_{B}T$.  At very low
refrigerator temperatures, FWHM is 
temperature independent, and the 
base electron temperature is $\sim$70\,mK. A magnetic field
of 1T is applied to supress superconductivity in the leads. 

In Fig.~2-d we show the line-shape of the level when the Al leads are
superconducting and when the superconductivity is suppressed with a
field of 1T applied parallel to the film plane. In zero field, the
line-shape traces the BCS density of states in the aluminum leads.
The shift of the level which occurs when the leads change from
superconducting to normal is proportional to the BCS gap in Al, and it
depends on the ratio of the
capacitances between the particle and the two leads.  The shifts
of the levels which we observe when Al is driven normal (not shown) 
indicate that all the levels have the same capacitances to the
leads, as expected if all resonances occur in a single grain.

Fig.~3 shows the excitation spectra ($dI/dV$) in three samples, in a
field of 1T to suppress the superconducting gap in the leads.  Peaks
occur at the energy levels of the particle. For each of the samples,
the corrective factor between voltage and particle energy is shown in
the figure. In sample 2, not all the levels correspond to the same
number of electrons on the particle.\cite{charge}

In the single-electron model, energy levels near the Fermi energy
should be distributed with only
small fluctuations between the successive level spacings (spectral 
rigidity).  The spectra
shown in Fig.~3 are very different from such single-electron
excitation spectra.  At low energies, spectroscopic peaks are sharp
and well separated.  As the excitation energy is increased, the
density of tunneling resonances increases, and the peaks have a 
tendency to cluster.
Eventually, we can not resolve individual
energy levels anymore; however we can still resolve envelopes (broad 
resonances) of 
many overlaping tunneling resonances. The broad resonances increase
in width as the energy increases, and finally, 
broad resonances merge forming a continuum 
in the
tunneling density of electronic states.

The increase in the density of tunneling resonances with energy
seen in Fig.~3 can be explained by electron-electron
interactions. In the non-equilibrium model due
to Agam {\it et al.}\cite{agam}, 
the same single electron
state can occur with different excited 
configurations of
the other electrons in the particle.
The theory of Altshuler {\it et  al.}\cite{altshuler} 
describes the delocalization (in Fock space) 
of a quasiparticle by
the creation of electron-hole pairs. This leads to an increase in the
density of discrete levels with increasing energy and to the 
existence of quasiparticles
with finite lifetime above a certain energy. The highest energy at
which the quasiparticles can be resolved is predicted 
to be the Thouless 
energy.\cite{sivan} 

The idea that
the Thouless energy sets the limit of the observability of the
discrete energy spectrum in an interacting electron system has been
first experimentally studied by Sivan {\it et al.}\cite{sivan}, 
on a disordered 
semiconducting quantum dot. In that work,
the authors' data analysis shows that there has to be a finite energy 
above 
which the
energy levels in a quantum dot can not be resolved.  
In our particles, unlike semiconducting quantum dots, 
the charging energy is so much larger than the level spacing 
that we can, for the first time, access 
and display the entire progression, 
out to the high energy horizon,
all within the
first step of Coulomb blockade. 
  
\begin{figure}
\epsfxsize=8.6cm
\epsfclipon
\epsffile{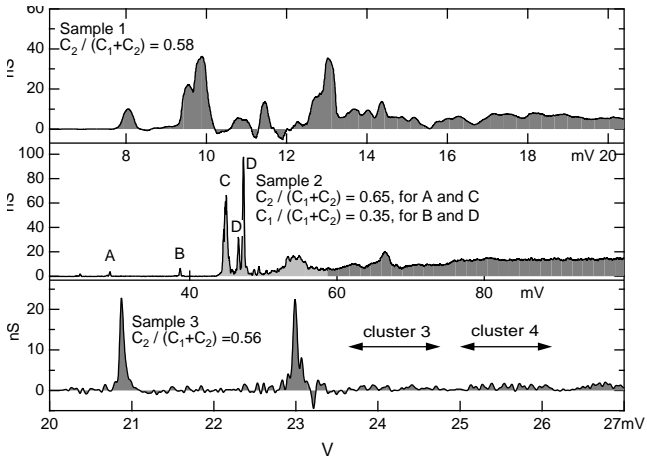}
\caption{Excitation spectra in 
three different samples, at $T=30mK$ and $H=1T$. In order to convert 
from voltages to particle eigenenergies, voltage must be 
multiplied by the appropriate correction factor, indicated for each 
graph.}
\label{figure3}
\end{figure}

Because the density of observed peaks increases rapidly with energy,
we rely on spectral rigidity and estimate the single-electron level
spacing $\delta$ as the distance between the two lowest energy
peaks. The agreement between the measured level spacing and the
calculated one (from the grain volume inferred from the capacitance)
is within a factor of 2, as shown in Table 1. We note
that in sample~3, the distance between the low energy clusters is
approximately the single-electron level spacing, consistent with the
theories of electron-electron interactions.

If we assume that the electronic motion in the
particle is ballistic, and that the surface scattering is diffusive,
then the mean-free path $l$ of an electron 
confined in the particle is approximately $D$. To diffuse through the 
entire particle volume however, an electron should scatter from the 
particle surface about three times. Thus, we estimate the
electronic traversal time as $3D/v_{F}$
(essentially, the factor of 3 is the number of dimensions).
The Thouless energy is given  approximately as $E_{T}=\hbar v_{F}/3D$ 
= 
37meV, 75meV and 40meV, for samples 1, 2, and 3, 
respectively. It can
be seen from samples 1 and 2 that these estimates of the Thouless 
energy
are consistent within a factor of 2 of the bias voltage at which the 
spectra 
become independent of energy. 

Note that the agreement between the theory and our data is better if
we compare the {\em bias} voltage with the Thouless energy, i. e., if
we do not correct the Thouless energy with capacitive division
prefactors.  Our explanation is that the highly excited states
predicted by the Agam model can initiate the formation of
electron-hole pairs.  The maximum number of a particle's excited
states which can form when electrons exit the particle, is given by
the bias voltage divided by the level spacing. This number is
$\sim$10, 12, and 20 for samples 1, 2, and 3,
respectively.\cite{steady} Thus, the role of large bias voltage (due
to large charging energy) is to reduce the range of energies where the
levels can be resolved. If $q_{0}\approx e/2$, so that there is little
Coulomb blockade, we would expect to see many more resolved levels, as
found by Ralph {\it et al.} in gated aluminum particles.

We have measured the evolution of the spectrum with applied magnetic
field. In a metallic particle, 
every orbital state has a 2-fold spin
degeneracy. Let $n$ be the number of electrons on the
particle after an electron has tunneled onto (or off) the particle. If
$n$ is even, the total spin in the ground state is zero in the
simplest model of weakly interacting electrons, and the first
tunneling resonance shifts but does not split with magnetic field. If
$n$ is odd, there is an unpaired spin, and all the levels split with
magnetic field. (Because the particle is small, the effect of the
field on electronic orbits is negligible.) 

Zeeman splittings in agreement with this 
simple model have been measured in 
Al nanoparticless\cite{ralph1} and some carbon
nanotube ropes\cite{Mceuen}. 
However, in ferromagnetic particles, and,
surprisingly, also in large semiconducting quantum dots and
some carbon 
nanotubes, the expected spin degeneracy
is {\it not} observed experimentally.\cite{cobalt}
In this work, we 
do measure Zeeman splitting in Au nanoparticles, 
confirming that spin degeneracy is commonly
observed in simple metals, as expected,
even though it is not 
observed in some other systems.

Fig.~4 is the gray-scale
image of the magnetic field dependence of the energy spectrum in
sample~1, at negative bias voltage, where the particle is further away
from equilibrium than for positive bias. The low energy levels split
with magnetic field, and the splitting is linear with field. Because
the lowest level splits into two, we conclude that the number of
electrons is odd. 
From the data, we obtain that the g-factor in this resonance 
is $g=0.28$. Sample~3 has similar splittings,
with a $g$-value of 0.44. In sample~2, the lowest two peaks do not
split implying that the number of electrons on the particle is
even. The splitting of the higher energy peaks in sample 2 could not
be resolved.  

A $g$-factor of $\sim 0.3$ in Au nanoparticle eigenenergies
is much smaller than 
the g-factor of $2.1$ determined by electron spin resonance 
measurements in bulk Au.\cite{cesr}
We explain the weak Zeeman splitting of Au nanoparticle 
eigenenergies 
by the strong spin-orbit (s-o) interaction in Au.
The explanation 
is corroborated by the work by Salinas {\it et 
al.}\cite{salinas},
where eigenenergies of superconducting 
Al nanoparticles doped with 4\% of Au 
had g-factors in range 0.5-0.8, instead of 2, which
is the g-factor in pure Al. 
The effect of s-o interaction on the single-electron 
eigenstates\cite{halp}  is to mix 
up-spin and down-spin polarizations, which leads to reduced g-factors.

The discrepancy between the g-factors determined by spin resonance 
experiments in
bulk Au\cite{cesr} and the g-factors of $\sim
0.3$ obtained here by tunneling spectroscopy
implies that 
for
a high-Z element such as Au, 
these two techniques measure two different 
quantities, which can not be directly compared.
Further theoretical study is needed to explain this discrepancy
and derive the g-factors in Au nanoparticles.  
We speculate that the nanoparticle geometry
(very large surface to volume ratio) plays an
essential role in 
g-factors reduction, pressumably
because the spin-flip scattering in Au nano-particles 
occurs at the particle surface. 

Note that in Fig. 4, resonance $\beta$ does not split
with magnetic field, and resonances $\gamma$ and $\gamma'$
split with magnetic field and cross; the analysis 
of these effects is beyond the scope of the
present paper.\cite{drago2}

\begin{figure}
\epsfxsize=7.6cm
\epsffile{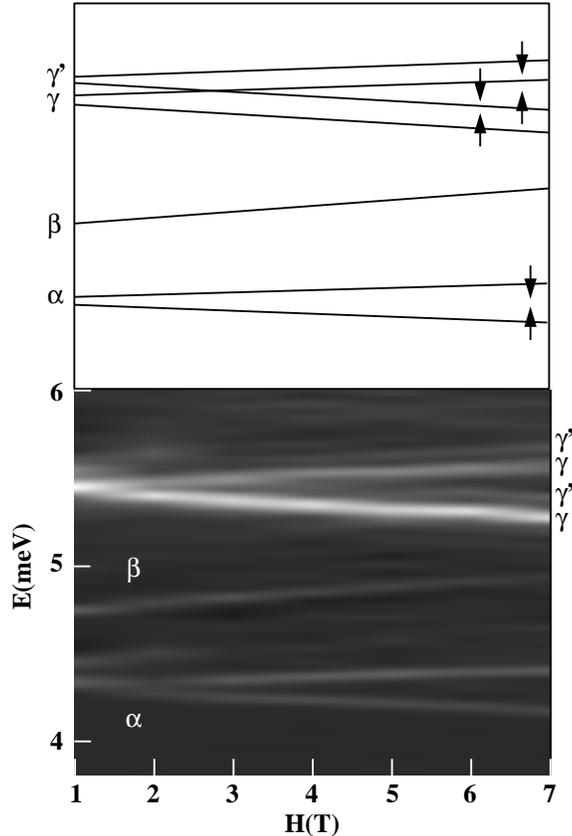}
\caption{Magnetic field dependence 
of the energy spectrum in sample 1.  Top schematic is a guide to the 
eye.}
\label{figure4}
\end{figure}

In conclusion, we have performed spectroscopy of discrete energy
levels in Au nanoparticles, with a new tunneling configuration.  
Data for
some of our samples include the entire spectrum starting from single
electron levels at low energy, progressing to a regime 
where only broad resonances (composed of many 
discrete levels) can be resolved, and ending in a continuum. We have
observed Zeeman splitting under applied magnetic field and found very
small $g$-factors, much smaller than the g-factors in bulk Au. 

The authors wish to thank Bertrand Halperin, Yuval Oreg, Piet Brouwer,
Dan Ralph, Steve Shepard and Sarah
Pohlen for useful discussions. This work has been supported in part 
by NSF
grants DMR-97-01487, DMR-94-00396, PHY-98-71810, and ONR grant
N00014-96-0108.

\end{multicols}

\widetext
\begin{table}
\caption{$C_{1}$, $C_{2}$, $R_1$, and $R_2$: junction capacitances 
and resistances determined from 
Coulomb staircase at 4.2K. D: particle base diameter estimated from 
$C_1+C_2$ assuming hemispherical shape. $\bar{\delta}$: 
Estimated spacing beween electron-in-a-box levels, based on particle 
volume. $\delta$: measured level spacing. $E_{T}$: The 
Thouless energy of particles, estimated as $\hbar v_{F}/3D'$, where 
$D'$
is the particle base diameter corresponding to {\it measured} level 
spacing $\delta$. $g$: determined from Zeeman splitting.}
\begin{tabular}{|c|c|c|c|c|c|c|c|c|c|c|}
	\hline	
Sample&$C_{1}[$aF$]$&$C_{2}[$aF$]$&$R_{1}+R_{2}[G\Omega$]&$R_{1}/R_{2}$&$D 
[nm]$& 
	$\bar{\delta}$ [meV]&$\delta$ [meV]&$E_{T} [meV]$&Parity&$g$\\
	\hline
	1 & 4 & 5.5 & 0.15 & 7.9 & 9 & 0.65 & 1 & 37 & ODD &0.28\\
	\hline
	2 & 0.9 & 1.67 & 0.066 & 2 & 4.7 & 4.6 & 7 & 75 & EVEN &-\\
	\hline
	3 & 1.9 & 2.4 & 1.25 & $>$2.5 & 6 & 2.1 & 1.2 & 40 & ODD & 0.45\\
	\hline
\end{tabular}
\end{table}


\begin{references}
\bibitem{charge}
We distinguish between the two sequences by 
measuring the normal-superconducting peak shifts, 
and by 
comparing the FWHM of the 
peaks as a function 
of temperature, both of which depend on the capacitance ratio.
Peaks A and C in sample 2 arise
from an electron tunneling onto the particle then tunneling off,
whereas for peaks B and D, the tunneling occurs in the opposite
order. In samples 1 and 3, the low energy peaks correspond to the same
charge state of the particle.
\bibitem{steady}
In a steady state, about 1/2 of the particle 
excitation energy is transferred to the drain electrode.
\end{references}
\end{document}